\documentclass[twocolumn]{aastex62}

\usepackage{graphicx}   
\usepackage{amsmath}    
\usepackage{amssymb}    
\usepackage{diagbox}    
\usepackage{subfigure}
\usepackage{float}
\usepackage{booktabs}

\received{January 1, 2018}
\revised{January 7, 2018}
\accepted{\today}
\submitjournal{ApJ}

\shorttitle{The formation of sdA}
\shortauthors{J. Yu et al.}

\begin{document}

\title{The formation of subdwarf A-type stars}

\correspondingauthor{Guoliang L\"{u}, Xuefei Chen}
\email{guolianglv@xao.ac.cn, cxf@ynao.ac.cn}

\author[0000-0002-0786-7307]{Jinlong Yu}
\email{yjl199502@163.com}
\affiliation{School of Physical Science and Technology, Xinjiang University, Urumqi, 830046, China}


\author{Zhenwei Li}
\affiliation{Yunnan Observatories, Chinese Academy of Sciences, Kunming, 650011, China}
\affiliation{Key Laboratory for the Structure and Evolution of Celestial Objects, Chinese Academy of Science}
\affiliation{University of the Chinese Academy of Science, Yuquan Road 19, Shijingshan Block, 100049, Beijing, China}
\nocollaboration

 \author{Chunhua Zhu}
\affiliation{School of Physical Science and Technology,
Xinjiang University, Urumqi, 830046, China}

\author{Zhaojun Wang}
\affil{School of Physical Science and Technology,
Xinjiang University, Urumqi, 830046, China}

\author{Helei Liu}
\affil{School of Physical Science and Technology,
Xinjiang University, Urumqi, 830046, China}

\author{Sufen guo}
\affil{School of Physical Science and Technology,
Xinjiang University, Urumqi, 830046, China}

\author{Zhanwen Han}
\affiliation{Yunnan Observatories, Chinese Academy of Sciences, Kunming, 650011, China}
\affiliation{Key Laboratory for the Structure and Evolution of Celestial Objects, Chinese Academy of Science}
\affiliation{Center for Astronomical Mega-Science, Chinese Academy of Science, 20A Datun Road, Chaoyang District, Beijing 100012, China}
\nocollaboration

\author{Xuefei Chen}
\affiliation{Yunnan Observatories, Chinese Academy of Sciences, Kunming, 650011, China}
\affiliation{Key Laboratory for the Structure and Evolution of Celestial Objects, Chinese Academy of Science}
\affiliation{Center for Astronomical Mega-Science, Chinese Academy of Science, 20A Datun Road, Chaoyang District, Beijing 100012, China}
\nocollaboration

\author{Guoliang L\"{u}}
\affil{School of Physical Science and Technology,
Xinjiang University, Urumqi, 830046, China}




\begin{abstract}
  Subdwarf A-type stars (sdAs) are objects that have hydrogen-rich spectra with surface gravity  similar to that of hot subdwarf stars but effective temperature below the zero-age horizontal branch (ZAHB). They are considered to be metal-poor main sequence (MS) stars or extremely low-mass white dwarfs  (ELM WDs). In this work, using the stellar evolution code, Modules for Experiments in Stellar Astrophysics (MESA), we investigate the sdAs formed both by the evolution of (pre-)ELM WDs in double degenerate systems (DDs) and metal-poor main sequence with the single evolution models. We find that both of the evolutionary tracks of ELM WDs and metal-poor MSs can explain the observation properties of sdAs. However, the proportion between these two populations are uncertain. In this work, we adopt the method of binary population synthesis of  both ELM WDs in the disk and metal-poor MS in the halo to obtain the populations of ELM WDs and metal-poor MSs at different stellar population ages and calculate their proportions. We find that the proportion of metal-poor MSs to sdA population for a stellar population of 10 Gyr is $\sim$98.5 percent, which is consistent with the conclusion that most sdAs ($>$95 percent) are metal-poor MSs. And the proportion of ELM WDs (metal-poor MSs) to sdA population increases (decreases) from 0.1\% (99.9\%) to 20\% (80\%)  with the stellar population ages from 5 to 13.7 Gyr.
\end{abstract}




\section{Introduction}
Using the data of the Sloan Digital Sky Survey (SDSS), \citet{Kepler2016}
found thousands of hydrogen-rich objects with effective temperature $T_{\rm eff}$ between about 7 000 K and 20000 K (most less than 10 000 K) below the zero-age horizontal branch (ZAHB). Their surface gravities in logarithmic log $g$ are about 4.5 to 6.0, between the main sequence (MS) stars and white dwarfs (WDs), and are therefore classified as subdwarf A-type stars (sdAs) \citep{Kepler2016}. There are two possible explanations for the nature of sdAs, that is, they are metal-poor A/F type MS stars or extremely low-mass white subdwarfs (ELM WDs).

  ELM WDs are generally helium WDs and have masses less than 0.3 $\rm M_\odot$ (log $g$ $\approx$ 6.0), with 8 000 K $\lesssim$ $T_{\rm{eff}}$ $\lesssim$ 22 000 K and 5.0 $<$ log $g$ $<$ 7.0 \citep{Brown2013}. It's a very interesting topic and has been discovered in recent years by several surveys such as  WASP, SDSS, ELM Suvery, and the Kepler mission \citep{Brown2010,Kilic2011a,Brown2012,Kilic2012b,Brown2013,Gianninas2015,Brown2016a}.
   ELM WDs are quite different from other WDs, since the evolution timescale for a single star to evolve into a helium-core WD would exceed the age of the Universe by far, \citet{Bergeron1992} concluded that they must be in close binaries, which was confirmed by observations (\citep{marsh1995a,Marsh1995b}, and others) thereafter. More than 100 (pre-) ELM WDs have been discovered and most are found by the ELM Survey led by Brown and Kilic \citep{brown2017a,brown2017b,Kilic2011a}. Meanwhile, most of the ELM WDs are formed in DDs. Observationally, (pre-)ELM WDs are found in binary systems with a companion star such as the A- or F-type dwarfs(EL CVn-type systems, \citep{maxted2011,maxted2013,maxted2014a}, related theoretical study see \citet{chen2017,Zhu2019} ), the millisecond pulsars \citep{Istrate2014a,Istrate2014b,Zhu2015,Istrate2016} or another (typically a carbon-oxygen) WD (such as those in ELM Survey, related theoretical study see \citet{li2019}). The ELM WDs give us important information about the evolution of close binaries and the gravitational foreground noise \citep{Kilic2011b}. According to binary evolution theory, they can only be formed by either stable Roche lobe overflow channel (RL channel) or common envelope ejection channel (CE channel), and have very short orbital periods.

  The properties of sdAs are very similar with those of (pre-) ELM WDs. Furthermore, by analyzing the data of Gaia release 2 (DR2), \citet{Pelisoli2019a} recently found 50 new high-probability (pre-)ELM WDs from the sdA samples with reliable Gaia parallaxes. Meanwhile, In the latest paper of \citep{Pelisoli2019b}, they present a catalogue of 5762 ELM candidates selected from the second data release of Gaia (DR2). Based on these observations, it will become helpful to detect and improve theoretical models of ELM WDs (\citealt{Corsico2014,Corsico2016,Istrate2016}).

 However, according to the radial velocities from SDSS subspectral, \citet{Hermes2017} found that the vast majority of sdAs are not in close binaries, and they then concluded that most sdAs (over 99 percent) are unlikely to be ELM WDs.
 Similarly, \citet{brown2017a} suggested sdAs\footnote{ They demonstrate that surface gravities derived from pure hydrogen models suffer a systematic $\sim$ 1 dex error for
 sdA stars, likely explained by metal line blanketing below 9000 K. Therefore, \citet{Pelisoli2019a} fit the spectra by atmosphere models with metals added in solar abundance. And we choose the later as the sdA samples in the paper.} are metal-poor A/F type MS stars in the halo, a small part of which have properties similar to that of ELM WDs. By analyzing 5 sdAs in eclipsing binaries in detail, \citet{brown2017a} further obtained the masses of the sdAs and their companions, i.e. 1.2 $\rm M_\odot$ and 0.8 $\rm M_\odot$, respectively. Recently, a series of study by \citep{Pelisoli2018a,Pelisoli2018b,Pelisoli2019a} confirm that most sdAs are metal-poor A/F type MS stars and the small part of them are ELM WDs. {\bfseries \citet{kepler2019} also state the majority of sdAs are metal-poor MSs and only a small part are (pre-)ELM WDs.}

 In this study, we aim to study the possibilities of sdAs as ELM WDs and metal-poor MS stars theoretically,
 by investigating the properties of ELM WDs and metal-poor MS stars and comparing with the observations.
 Considering that most of the ELM WDs are formed in DDs, we mainly evolve ELM WDs in DDs.
 For the binary population synthesis (BPS) of ELM WDs, we consider that most of them are in the Galactic disk. By the Gaia DR2 data, \citet{Pelisoli2019a} and \citet{kepler2019} obtained the parallax, and could estimate the distances, luminosities and radii of sdAs. They found that most sdAs are metal-poor MSs in the Galactic halo. Based on the observational research of \citet{Bland-Hawthorn2016} and \citet{Deokkeun2013}, we set the metallicity $Z$ = 0.02 in the disk and the metallicity $Z$ = 0.0001 in the halo.
 By the method of binary population synthesis (BPS) of both ELM WDs and metal-poor MS, we estimate their proportions.
 In \S 2 we present model for the formation and evolution of ELM WDs and metal-poor MSs. In \S 3, the result calculated by a method of BPS is given. In \S 4, the main conclusions are shown.

\section{MODEL}
 As the last section mentioned, sdAs are composed of two populations: ELM WDs and metal-poor MSs.
 Therefore, we evolve the binary evolution models to obtain the ELM WDs in DDs and
 the single evolution models to obtain the metal-poor MS.
\begin{figure*}
\begin{tabular}{lr}
\includegraphics[width=0.5\textwidth]{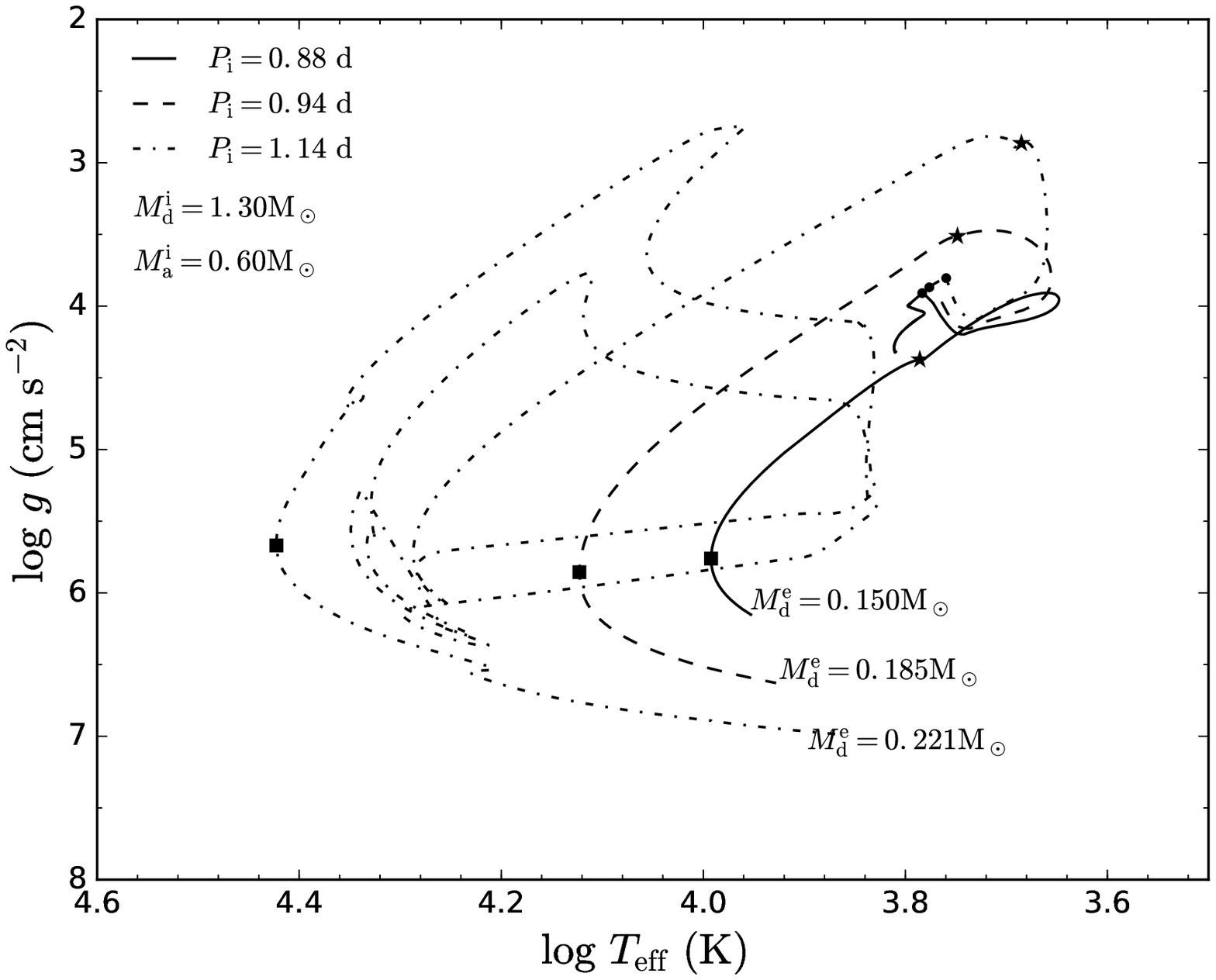}
\includegraphics[width=0.5\textwidth]{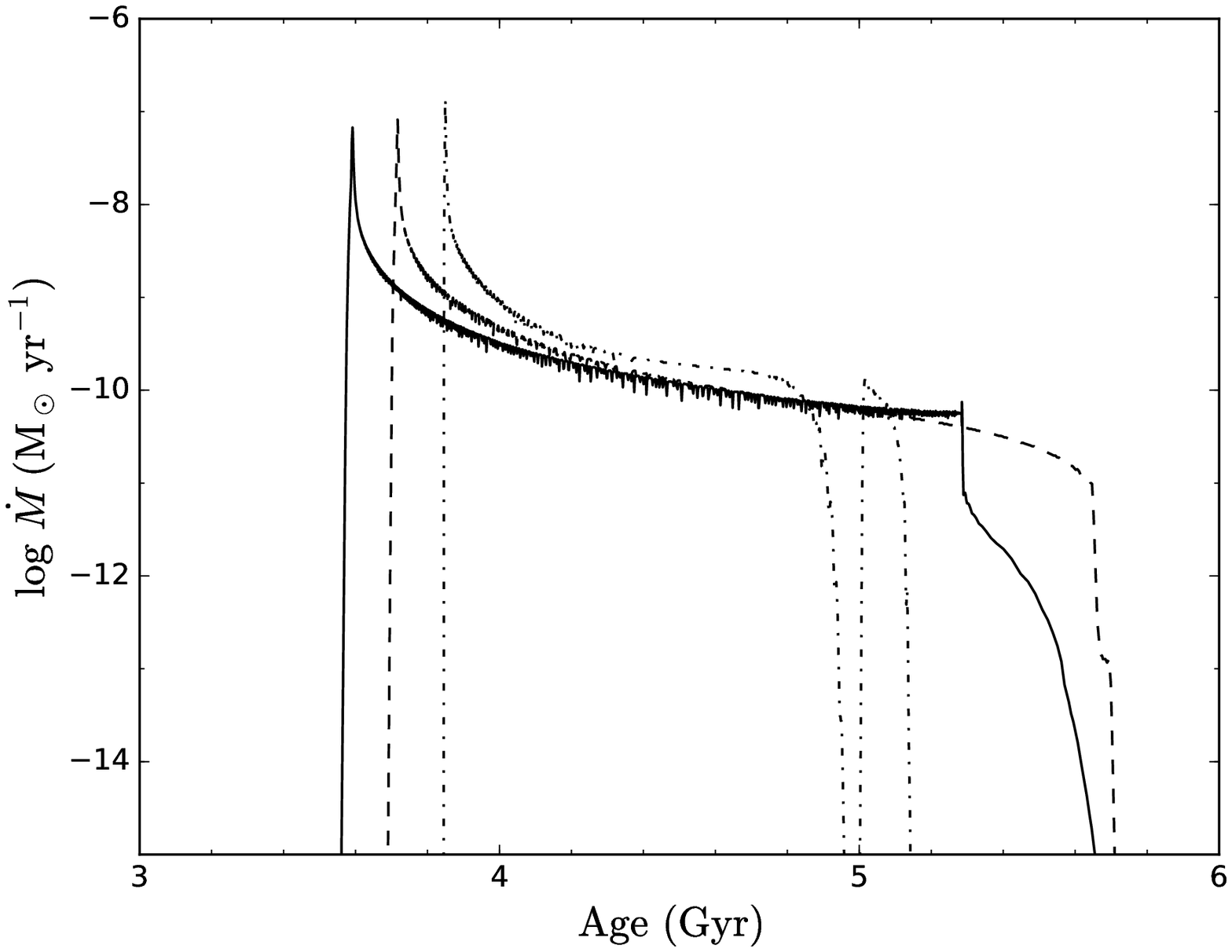}\\
\end{tabular}
    \caption{ Three typical evolutionary tracks of ELM WDs in the $T_{\rm{eff}}-\log g$ plane (left panel) and the mass-transfer rate changes  (right panel) with stellar age.
    The binary systems have the same initial component masses (the initial mass of the donor, $M_{\rm{d}}^{\rm i}$ = 1.3 $\rm M_{\odot}$;
    the initial mass of accretor, $M_{\rm{a}}^{\rm i}$ = 0.6 $\rm M_{\odot}$) but different initial orbital periods. The solid, dashed, dash-dot lines represent initial orbital period $P_{\rm i}$ = 0.88 , 0.94 , 1.14 day, respectively. The circles, stars and squares represent the beginning of mass transfer (MT), the end of MT and the maximum temperature, respectively.
  The final masses of donors, $M_{\rm{d}}^{\rm e}$, are 0.15, 0.185, 0.221 $\rm M_{\odot}$, which are showed at the end of the three typical evolutionary tracks, respectively.}
\label{fig1}
\end{figure*}

\subsection{Initial parameters and the simulation code}
In order to obtain the ELM WDs in DDs, we use the binary evolution models. According to the theory of binary evolution, there are two main evolution channels that lead to the formation of ELM WDs: the stable Roche lobe overflow (RLOF) channel and the common envelope (CE) ejection channel.
In this paper, we take advantage of the detailed binary evolution code Modules for Experiments in Stellar Astrophysics (\texttt{MESA}, version 8845, \citealt{paxton2011, paxton2013, paxton2015}) to carry out binary evolution calculations.  \citet{li2019} had produced several grids of evolution models of ELM WDs in DDs.
  Following \citet{li2019}'s work, we set the metallicity $Z$ = 0.02 and the hydrogen mass faction X= 0.70, the mixing length parameter is set to be MLT = 1.9. The mass transfer (MT) rate is given by \citet{Ritter1988} as follows:
\begin{equation}
    \dot{M}_{\rm{RLOF}}\propto \frac{R^3_{\rm{RL,d}}}{GM_{\rm{d}}}\exp\left(\frac{R_{\rm{d}}-R_{\rm{RL,d}}}{H_{\rm{P}}}\right),
    \label{eq:1}
\end{equation}
where $R_{\rm{d}}$ and $R_{\rm{RL,d}}$ are the stellar radii of the donor and its Roche lobe, respectively. $M_{\rm{d}}$ is the mass of donor, $H_{\rm{P}}$ is the pressure scale height of the atmosphere, and $G$ is the gravitational constant.

 {\bfseries We evolve these binary systems that consist of MS + CO WD, where the MS is donor and the CO \footnote {{\bfseries In this paper, because almost all the companions of ELM WDs in DDs in the ELM WD Survey are CO WDs, we choose CO WD as the accretor.}} is accretor.} The initial parameter space we selected is similar to \citet{li2019}:
 The initial donor mass ($M_{\rm{d}}^{\rm i}$ in $\rm M_{\odot}$ ) is 1.0, 1.1, 1.2,$...$, 2.0 (by steps of 0.1 $\rm M_{\odot}$);
 the initial accretor mass ($M_{\rm{a}}^{\rm i}$ in $\rm M_{\odot}$) is 0.5, 0.6, 0.7,$...$, 1.1 (by steps of 0.1 $\rm M_{\odot}$). The $M_{\rm{a}}^{\rm i}$ = 0.45 $\rm M_{\odot}$, which is usually considered the lowest mass of CO WD, is also be considered.
 The initial orbital periods we choose are slightly smaller than theirs in order to produce lower mass (see \citet{li2019} for more details).
  With this method, we produced the ELM WDs in DDs with masses from 0.105 $\rm M_{\odot}$ to 0.335 $\rm M_{\odot}$ and metallicity $Z$ = 0.02 in our calculation. The evolution will stop when evolutionary age reaches 13.7 Gyr. All of these ELM WDs are produced by RL channel.

 Similarly, in order to obtain the metal-poor MS, we evolve the single evolution models by MESA. In our simulation, the metallicity is taken as $Z$ = 0.01, 0.001, 0.0001, respectively. The opacity table is Type 2 opacities \citep{iglesias1996}, the mixing length parameter ($\alpha_{\rm{MLT}}$) is taken as 1.9, and the Ledoux criterion is used for convection. For each metallicity, we produced 4 metal-poor MSs with masses from 0.8 $\rm M_{\odot}$ to 2.0 $\rm M_{\odot}$ by steps of 0.4 $\rm M_{\odot}$. The evolution will stop when the stars reach giant branch (GB).

\subsection{ Evolution results }
\subsubsection{Binary evolution results for (pre-) ELM WDs}
 To better understand our model and analyze the evolution process, we present three typical evolutionary tracks of ELM WDs in the $T_{\rm{eff}}-\log g$ plane (the left panel of Figure \ref{fig1}) and MT rate changes with star age (the right panel of Figure \ref{fig1}). These binary systems have the same initial component masses ($M_{\rm{d}}^{\rm i}$ = 1.3 $\rm M_{\odot}$, $M_{\rm{a}}^{\rm i}$ = 0.6 $\rm M_{\odot}$) but different initial orbital periods: $P_{\rm i}$ = 0.88, 0.94 and 1.14 day.

 The donors in the system with shorter initial orbital periods fill Roche lobe earlier. For the binary system with $P_{\rm i}$ = 0.88 day,
 the donor fills its Roche lobe before the end of MS, and MT occurs. As the right panel of Figure \ref{fig1} shows, MT
 lasts $\sim$ 2 Gyr, and the donor lost most H-rich envelope (about 1.15 $\rm M_{\odot}$). Finally, it directly evolves into a (pre-) ELM WD with the mass of 0.15 $\rm M_\odot$ and reaches the maximum temperature.
 Compared with the former, the donor in the binary system with $P_{\rm i}$ = 0.94 day fills late its Roche lobe, and
 it has larger core. From the right panel of Figure \ref{fig1}, we can see the MT rate is slightly larger than the above one. The MT lasts  $\sim$ 1.9 Gyr, and the donor lost most H-rich envelope (about 1.115 $\rm M_{\odot}$). Finally, it directly evolves into a (pre-) ELM WD with the mass of 0.185 $\rm M_\odot$.
 However, for the binary system with $P_{\rm i}$ = 1.14 day, the MT is different from the first two cases significantly. The donor fills latest its Roche lobe. The MT occurs at the end of MS and MT rate is the biggest, as the right panel of Figure \ref{fig1} shows. When the MT lasts $\sim$ 1 Gyr, the first dredge-up occurs. The internal helium element mixes with the hydrogen-envelope and the hydrogen abundance decreases in envelope. This process causes a decrease in the nuclear reaction rate. So the radius of the donor shrinks into the radius of Roche lobe and MT rate decreases significantly. When the mass of the helium core increases enough, the hydrogen envelope burns again and the donor fills Roche lobe again. So the MT rate increases again, and MT ends until the most envelope is lost (about 1.07 $\rm M_{\odot}$). Then it evolves into a (pre-) ELM WD with a massive core (0.221 $\rm M_{\odot}$). Finally, the donor suffers several strong H-shell flashes before it evolves into the cooling phase.
\begin{figure}
  \centering
  \includegraphics[width=0.5\textwidth]{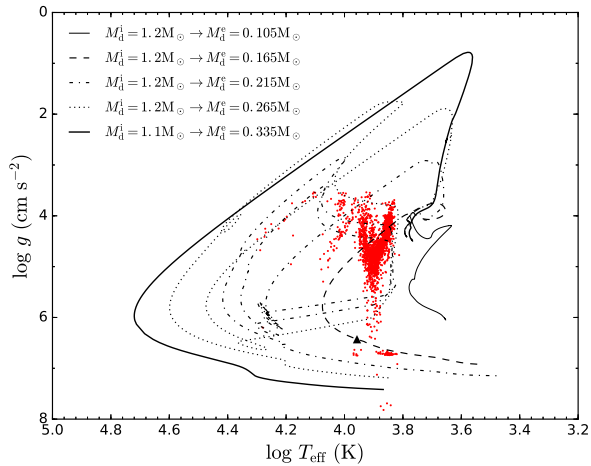}
  \caption{Evolutionary tracks of the donors in the $T_{\rm{eff}}-\log g$ plane from the zero-age main sequence to the maximum age. These red dots are sdAs samples obtained from the SDSS data release 12 (DR12).
  In order to cover all samples, we extend the evolution time to 20Gyr. The black triangle represents the age of 13.7Gyr.   }
  \label{fig2}
\end{figure}


\begin{figure*}
\begin{tabular}{lr}
\includegraphics[width=0.5\textwidth]{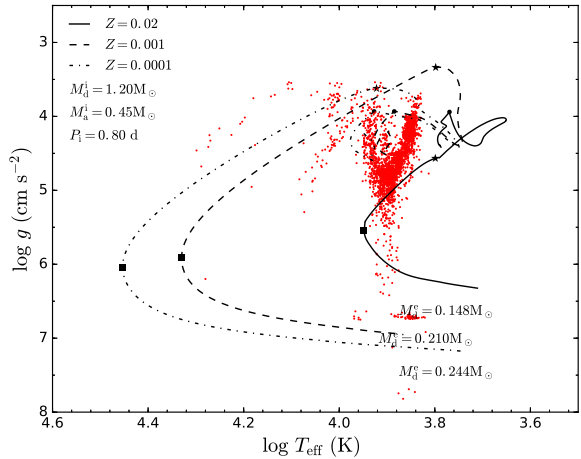}
\includegraphics[width=0.5\textwidth]{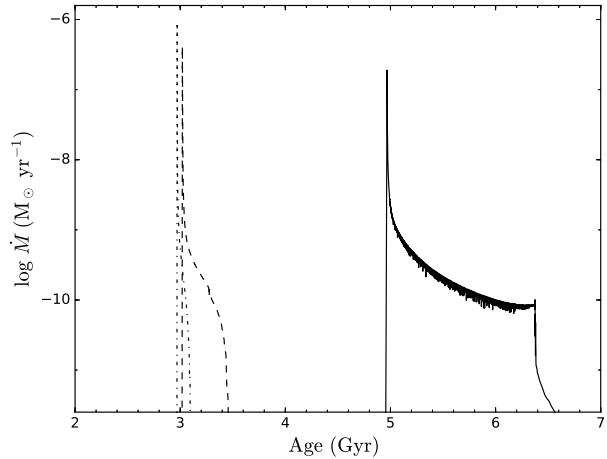}\\
\end{tabular}
\caption{ Similar to Figure \ref{fig1} but with different metallicity, where the binaries have the same initial componet masses ($M_{\rm{d}}^{\rm i}$ = 1.2 $\rm M_{\odot}$, $M_{\rm{a}}^{\rm i}$ = 0.45 $\rm M_{\odot}$) and same initial orbital period $P_{\rm i}$ = 0.80 day. The solid, dashed, dash-dot lines represent metallicity $Z$ = 0.02, 0.001, 0.0001, respectively. The final masses of donors, $M_{\rm{d}}^{\rm e}$, are 0.148, 0.210, 0.244 $\rm M_{\odot}$, respectively.
The circles, stars and squares are same as the markers in Figure \ref{fig1}.}
\label{fig3}
\end{figure*}

 Figure \ref{fig2} shows the comparison of evolution tracks with observations. We selected  5 typical evolutionary models: 0.105, 0.165, 0.215, 0.265, 0.335 $\rm M_{\odot}$, including the edges of the mass range. The sdAs samples obtained from the SDSS data release 12 (DR12) are marked with red dots. These typical evolutionary tracks cover the whole sdAs samples very well and they fully reflect that different parameter spaces lead to different evolution tracks. The ELM WDs with the masses between 0.215 $\rm M_{\odot}$ and 0.315 $\rm M_{\odot}$ suffer several strong H-shell flashes in the H-rich envelope and then evolve into the cooling phase.
 For these sdAs, if we only consider the values of $T_{\rm{eff}}$ and $\log g$, the majority of sdAs could be explained as (pre-) ELM WDs, and the rest may be ELM WDs in cooling stage.
 However, in \citet{Pelisoli2019a}, they obtain the parallax of the sdAs by using Gaia DR2 data, and thus obtain the distance, luminosity and radius, which is an important parameter to distinguish between the metal-poor MSs and the ELM WDs. We simply present our evolutionary tracks in the $T_{\rm{eff}}-\log g$ diagram and compare them with the observations here. In section \S 3.2, we study the radius distribution by the BPS in detail.

 The metallicity $Z$ can affect the stellar structure and evolution. Here, we present the evolutionary tracks of ELM WDs in Figure \ref{fig3} with different metallicity, which is $Z$ = 0.02, 0.001 and 0.0001. The three binaries have the same component masses, that is, $M_{\rm{d}}^{\rm i}$ = 1.2 $\rm M_{\odot}$, $M_{\rm{a}}^{\rm i}$ = 0.45 $\rm M_{\odot}$ and the same initial periods, $P_{\rm i}$ = 0.80 day.
  The stars with low metallicity have a lower opacity, then a smaller radiation pressure, therefore, the density and temperature in the center of the star are higher, and the stars have larger core and the burning efficiency of hydrogen is higher.
 They evolve quickly, and fill Roche lobe earlier. For the $Z$ = 0.0001, as the right panel of Figure \ref{fig3} shows,
 the donor fills its Roche lobe earliest and the MT occurs when its core mass is about  0.135 $\rm M_{\odot}$. The MT lasts $\sim$ 0.1 Gyr,
 and the core mass increases to 0.244 $\rm M_\odot$ and most H-rich envelope (about 0.95 $\rm M_{\odot}$) transfers into its companion. Finally, it directly evolves into
 (pre-) ELM WD. For the $Z$ = 0.001, it is similar to the former, but the donor fills late its Roche lobe and has a smaller core (about 0.11 $\rm M_{\odot}$). Therefore, the MT lasts for a longer time ($\sim$ 0.42 Gyr). When the MT ends, the donor loses most H-rich envelope (about 0.99 $\rm M_{\odot}$). Finally, it directly evolves into
 (pre-) ELM WD with the mass of 0.210 $\rm M_\odot$. For the $Z$ = 0.02, from the right panel of Figure \ref{fig3}, we can see that the donor fills latest its Roche lobe. It has the smallest core (about 0.08 $\rm M_{\odot}$) when it leaves MS, therefore, the center temperature of the donor is lower and the radius expands more slowly, and the MT lasts for a longer time ($\sim$ 1.6 Gyr). When the MT ends, the donor loses most H-rich envelope (about 1.05 $\rm M_{\odot}$). Then it evolves into (pre-) ELM WD with the mass of (0.148 $\rm M_{\odot}$).
 Figure \ref{fig3} illustrates that the evolutionary tracks of ELM WDs with low metallicity also agree quite well with some samples. Therefore, a proportion of sdAs (i.e.the objects with high temperature and gravity) are likely ELM WDs with low metallicity.

\subsubsection{Single evolution results for metal-poor MS}
\begin{figure}
  \centering
  \includegraphics[width=0.5\textwidth]{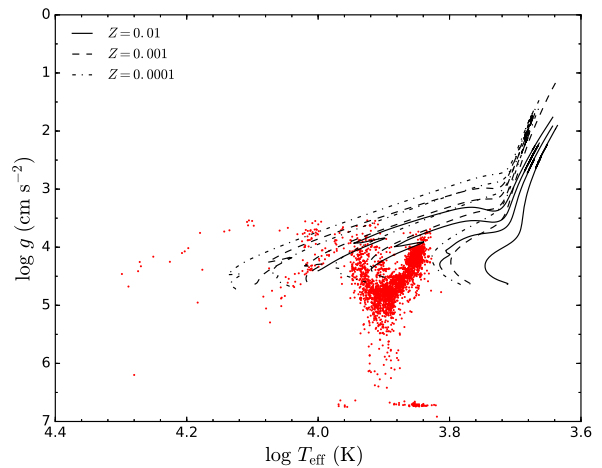}
  \caption{Evolutionary tracks of the single evolution models with different metallicity in the $T_{\rm{eff}}-\log g$ plane from the ZAMS to the GB. The solid, dashed, dash-dot lines represent metallicity $Z$ = 0.01, 0.001, 0.0001, respectively. From the right to the left the star masses are 0.8, 1.2, 1.6 and 2.0 $\rm M_{\odot}$, respectively. The red dots are the same as the dots shown in Figure \ref{fig2}.}
  \label{fig4}
\end{figure}

 The vast majority of sdAs may be the metal-poor MS stars \citep{Pelisoli2018a,kepler2019}.
 Figure \ref{fig4} shows the evolutionary tracks of single evolution models from ZAMS to GB with different metallicity in the $T_{\rm{eff}}-\log g$ diagram. As showed in Figure \ref{fig4}, these evolution tracks can only cover the upper part of the sample derived from solar abundance spectral fitting. As stated in \citet{Pelisoli2018a}, these objects cannot be explained as simply metal-poor MS of A-F type by analysing the estimated distances and spacial velocities.
 As mentioned above, It is necessary to study the distribution of the radius and compare it with the observations. We simply present our evolutionary tracks in the $T_{\rm{eff}}-\log g$ diagram compared with the observations here. In section \S 3.2, we show our radius distribution obtained by the BPS and compared them with the observations in detail.
\section{BINARY POPULATION SYNTHESIS}
 We perform a Monte Carlo simulation to obtain the populations of ELM WDs in DDs in the disk and metal-poor MSs in the halo, respectively.

\subsection{Basic assumption of BPS for ELM WDs in DDs and metal-poor MSs in the halo}
In order to show the distribution of the ELM WDs and metal-poor MS in the sdAs, we use the method of binary population synthesis (BPS).
Following \cite{Lu2012}, \cite{Zhu2015}, \cite{Lu2017}, \cite{Zhu2017} and \cite{Wang2018}, we combined rapid binary star evolution code BSE \citep{Hurley2000, Hurley2002} and the calculation results from MESA to get the whole ELM WDs populations. As introduced above, we use MESA to obtain the populations of ELM WDs and the parameter space produced by RL channel. Meanwhile, we use BSE to obtain the populations of ELM WDs produced by CE channel. The detailed method we employed are the same with \citet{li2019}, but for the CE channal, we set $\alpha_{\rm CE}$ = 1.0. Usually, the larger $\alpha_{\rm CE}$ is, the more easily CE is ejected. It means that binary model with a large $\alpha_{\rm CE}$ can make more close binary systems survived after CE evolution while it with a small $\alpha_{\rm CE}$ can result in more merger of binary systems.
\citet{li2019} took $\alpha_{\rm CE}$ = 0.25, 0.5 and 1.0 in the different models, respectively. In their work, for larger $\alpha_{\rm CE}$, the proportion of ELM WDs from the CE channel increases, since more orbital energy is released to eject the CE. They selected $\alpha_{\rm CE}$ = 1.0 as the standard model. Therefore, for simplicity, in our work, we set $\alpha_{\rm CE}$ = 1.0.
 At first, in order to obtain the stellar population of CO WD+donor systems at the beginning of MT, we used a Monte Carlo simulation with 5$\times$ $10^{6}$ primordial binary systems in BSE. These systems can form ELM WDs either by stable mass transfer or by CE channel. Then, we interpolate in the grid from MESA to obtain the ELM WDs produced from stable mass transfer. If the parameters are included in the grid, we consider that these systems of ELM WDs were produced from RL channel. Conversely, if the parameters are not included in the grid, the systems either produce by CE channel or cannot produce ELM WDs.

 To obtain the populations of ELM WDs produced by CE channel, we choose the remained systems go through the CE process by BSE. By analyzing the evolutionary results, we retain these systems that can produce the ELM WDs. At last, we could obtain the whole population of ELM WDs by combining the results of the RL channel and the CE channel. Based on \citet{Bland-Hawthorn2016} and \cite{Deokkeun2013}, we set the metallicity $Z$ = 0.02 in the Galactic disk. We assume that the constant star formation rate (SFR) is 2 $\rm M_{\odot}\;\rm yr^{-1}$ over the last 13.7 Gyr for the Galaxy \citep{chomiuk2011}.

 In our Monte Carlo simulation, we assumed that all stars are members of binaries. The circular orbits are also assumed. There are three main input parameters for BPS study: the initial mass function (IMF), the mass-ratio distribution and the initial orbital separation distribution. The mass distribution of initial primary follows the IMF of \citet{Miller1979} and is generated according to the formula of \citet{Eggleton1989}:
 \begin{equation}
      M = \frac{0.19X}{(1-X)^{0.75}+0.032(1-X)^{0.25}},
\label{eq:2}
\end{equation}
where $X$ is a random number that is uniformly distributed over a range from 0 to 1, meanwhile, the mass range of primary from 0.08 to 100 $\rm M_{\odot}$ is given. Mass ratio has a very important impact on the evolution of binary stars, in our work, we calculate the secondary mass that obtained from a constant mass ratio distribution \citep{Mazeh1992}, i.e. $n(q^{'})$ = 1, 0 $\leq$ $q^{'}$ $\leq$ 1, where $q^{'}$ = $M_{\rm d}$/$M_{\rm a}$. The distribution of initial orbital separation is taken to be uniform distribution in log $a$ for wide binaries. We use the distribution given by \citet{han1998} for compact binaries:
\begin{equation}
	an(a)=
	\begin{cases}
	0.07(a/a_0)^{1.2},\qquad a\le a_0 \\
	0.07,\qquad \qquad a_0 \le a \le a_1,
	\end{cases}
	\label{eq:3}
\end{equation}
where $a_0=10\;\rm{R}_{\odot}$, $a_1=5.75\times 10^6\;\rm{R}_{\odot}$ = 0.13 $\rm pc$. When $a>10\;\rm{R}_{\odot}$, the distribution is a constant of wide binary systems per logarithmic interval and results in approximately half of the binary systems having an orbital period less than 100 years\citep{chen2009,Lu2011,Lu2012}.

Similarly, we use employ BSE to obtain the population of metal-poor MS in the halo. We used a Monte Carlo simulation with 5$\times$ $10^{6}$ primordial binary systems in BSE to obtain MS + other star systems. The three main input parameters for BPS study: the initial mass function (IMF), the mass-ratio distribution and the initial orbital separation distribution, are same as above. {\bfseries The difference is that a single burst of total mass of $1 \times 10^{9}$ $\rm M_{\odot}$ is adopted\citep{Yu2010} and the metallicity ($Z$) is set to 0.0001 in the Galactic halo \citep{Bland-Hawthorn2016,Deokkeun2013}.}
For the ELM WDs and metal-poor MSs stars, in order to be better agreed with the observations, we obtain them at different stellar population ages: 5Gyr, 8Gyr, 10Gyr, 12Gyr and 13.7Gyr.
Because the number is affected by Poisson noise, we calculate the relative error of the numbers of ELM WDs in
DDs at different stellar population ages, which is lower than $1.3 \times 10 ^{-2}$. Similarly, for the metal-poor MSs, the Poisson noise is lower than $9.35 \times 10 ^{-3}$.

 Finally, we combine the two populations to get the results and compare them with the observations of sdAs in the $T_{\rm{eff}}-\log L$ plane and $T_{\rm{eff}}-\log R$ plane. Meanwhile, we count the proportion of ELM WDs and metal-poor MSs at different stellar population ages.

\subsection{The result of BPS}

 Considering that the brighter stars are easier to be observed, even if short lived, we should make a volume correction.
 Obviously, the theoretical number of (pre-) ELM WDs obtained from our binary models in the nearly constant-L contraction phase is less than the number in the cooling phase. But the objects in the nearly constant-L contraction phase are more brighter and more easily observed. Therefore, in order to take into account this fact, we adopted a volume correction for both ELM WDs and metal-poor MSs populations to estimate the numbers. We calculated the volumes $V_{\rm obs}$ of every model according to the formula below:
\begin{equation}
 V_{\rm obs} = a \times L^{3/2}
\label{eq:4}
\end{equation}
 where $L$ is the luminosity of the star, $a$ is a constant coefficient, which does not affect the final distribution. Firstly, according to Eq. (\ref{eq:4}), we have obtained the observable volume $V_{\rm obs}$ of each model. Then, we summed the observable volumes $V_{\rm obs}$ of all models in each bin and get the observable volumes $V_{a}$ in each bin. For each bin, we calculated the ratio of observable volumes to the minimum volumes $V_{min}$, and obtained the relative proportion of each bin by formula $p_{\rm obs}$ = $V_{a}$/$V_{min}$.

 \begin{figure}
  \centering
  \includegraphics[width=0.5\textwidth]{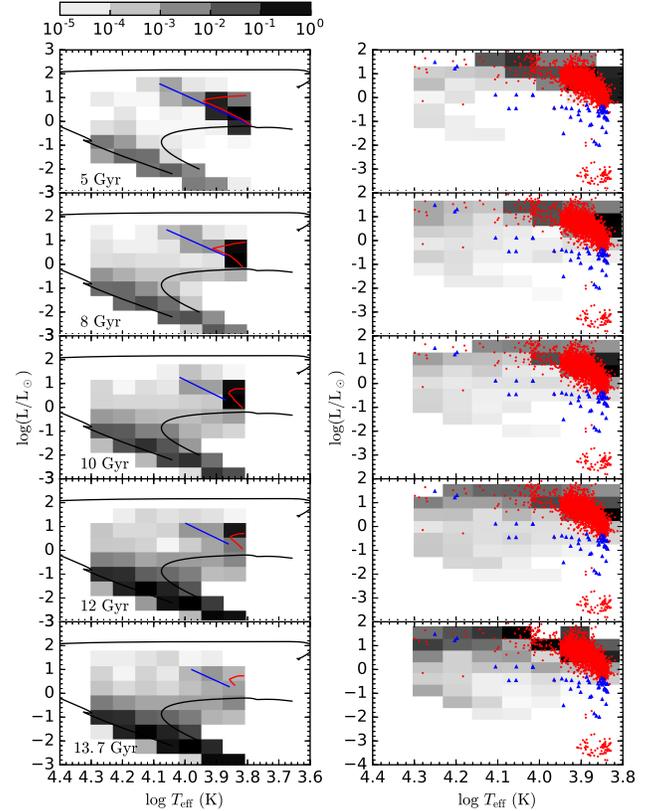}
  \caption{The distribution of BPS results of both ELM WDs in the Galactic disk (a constant star formation rate over the Galaxy age) and metal-poor MS in the halo (a single burst) at different stellar population ages in the $T_{\rm{eff}}-\log L$ plane. The left plane without a volume correction and the right plane with a volume correction. The red and blue solid lines represent two different systems of metal-poor MS population (MS + MS and MS + He/CO WD), respectively. The two evolutionary tracks of ELM WDs are showed with black solid lines. {\bfseries The final masses are 0.165 (lower black solid line) and 0.305 $\rm M_{\odot}$ (upper black solid line), respectively.} The red dots represent the sdAs samples. The 50 high-probability new (pre-) ELM WDs obtained from \citet{Pelisoli2019a} are marked by blue triangles.}
  \label{fig5}
\end{figure}

 Figure \ref{fig5} shows the distribution of BPS results of both ELM WDs in the disk (a constant star formation rate over the Galaxy age) and metal-poor MS in the halo (a single burst) at different stellar population ages in the $T_{\rm{eff}}-\log L$ plane. The left plane without a volume correction and the right plane with a volume correction. The red dots represent the sdAs samples. The 50 high-probability new (pre-) ELM WDs obtained from \citet{Pelisoli2019a} are marked by blue triangles. From left plane (without volume correction) of Figure \ref{fig5},
 we can see two distinct parts in each picture, one of which (upper right) is the distribution of metal-poor MS and the other part (lower left) is the distribution of ELM WDs. For the BPS results of metal-poor MS (upper right), there are two parts, one of which is the metal-poor MS + He/CO WD (blue solid line) and the other is the metal-poor MS + metal-poor MS (red solid line) with a lager orbital separation, which can be considered as the single star evolution.

\startlongtable
\begin{deluxetable}{ccccccc}
\tablecaption{The proportions of two populations at different stellar population ages \label{tab:1}}
\tablehead{
\colhead{Age (Gyr)} &
\colhead{5} &
\colhead{8} &
\colhead{10} &
\colhead{12} &
\colhead{13.7} &
}
\startdata
ELM WD & $0.1\%$ & $1\%$ & $1.5\%$ & $7\%$ & $20\%$ \\
metal-poor MS & $99.9\%$ & $99\%$ & $98.5\%$ & $93\%$ & $80\%$ \\
\enddata
\end{deluxetable}

 We can found that the proportion of ELM WDs increases and the proportion of metal-poor MSs decreases with the stellar population age, as shown in the left plane of Figure \ref{fig5}.
 When we consider the volume correction, the proportion of faint objects (most ELM WDs in the cooling phase) will decreased significantly, as shown in the right plane of Figure \ref{fig5}, and the corresponding proportions are presented in Table \ref{tab:1}. The proportions of metal-poor MSs for a stellar population of 10 Gyr is $\sim$ 98.5 percent, which is consistent with the conclusion that most sdAs ($>$ 95 percent) are metal-poor MSs. The metal-poor MSs dominate the proportion before 12 Gyr. On the other hand, these metal-poor MSs in the halo may be born after two billion years of the formation of the universe.
  For the samples ($\sim$ 40 stars) with lower luminosity that located in the lower right corner in each picture, they are consistent with the single evolution models of canonical mass WDs (see Figure 8 of \citet{Pelisoli2019a}) and are included in the Gaia white dwarf catalogue.

\begin{figure}
  \centering
  \includegraphics[width=0.5\textwidth]{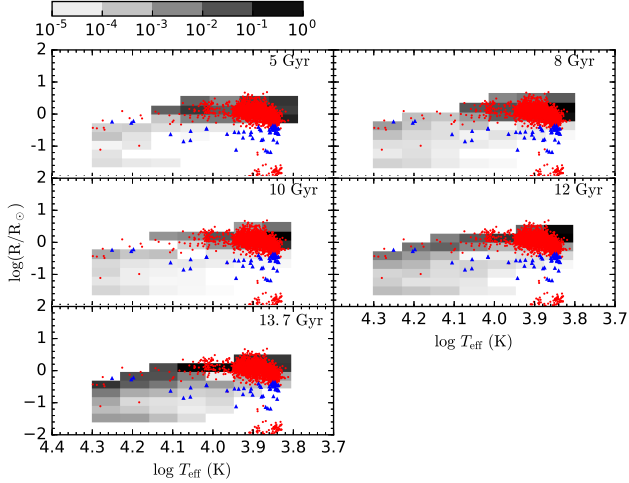}
  \caption{The distribution of radius $R$ and the comparison with the observations. The red dots represent the sdAs observation samples. The 50 high-probability new (pre-) ELM WDs obtained from \citet{Pelisoli2019a} are marked by blue triangles.}
  \label{fig6}
\end{figure}

 In the paper of \citet{Pelisoli2019a}, they first calculated the distance with the parallax obtained from the Gaia data, and then got the radius of the stars. This is an important parameter that distinguishes between (pre-)ELM WDs and metal-poor MSs. In order to compare to the observations, we present the distributions of radius at different stellar population ages in Figure \ref{fig6}. The red dots represents the sdAs observation samples. From this figure, we can found that the distributions of radius are consistent with observations before 12 Gyr, because the proportion of ELM WDs increases and the proportion of metal-poor MSs decreases with the stellar population age, especially after 12 Gyr.
  Meanwhile, for the samples ($\sim$ 40 stars mentioned above) with smaller radius that located in the lower right corner in each picture, they are inconsistent with the BPS result. Therefore, they are more likely to be canonical mass WDs by the single evolution.

\section{CONCLUSIONS}
  In this paper, we simulate the binary evolution models to obtain (pre-) ELM WDs in DDs and the single evolution models to obtain metal-poor MS stars with MESA. We present our evolutionary tracks in the $T_{\rm{eff}}-\log g$ plane and compare to observations. Based on the binary evolution models and single evolution models, we found sdAs may be composed of these two populations.
   By the method of binary population synthesis (BPS) of both ELM WDs in the disk (a constant star formation rate over the Galaxy age) and metal- poor MS in the halo (a single burst), we obtain ELM WDs and metal-poor MSs at different stellar population ages(5Gyr, 8Gyr, 10Gyr, 12Gyr and 13.7Gyr) and calculate their proportions. We found that the proportion of ELM WDs increases with the stellar population ages, but the proportion of metal-poor MSs will decreases. The proportions of metal-poor MSs for a stellar population of 10 Gyr is $\sim$ 98.5 percent, which is consistent with the conclusion that most sdAs ($>$ 95 percent) are metal-poor MSs. The metal-poor MSs dominate the proportion before 12 Gyr. Finally, we also compare the radius with observations and found the distribution of the radius is consistent with the observations before 12 Gyr. These metal-poor MSs in the halo may be born after many million years of the formation of the universe.

\section*{Acknowledgements}
This work received the generous support of the  National Natural Science Foundation of China,
project Nos. 11863005, 11763007, 11503008, 11733008 and 11521303 and the National Ten-thousand talents program and Yunnan province (No. 2017HCO18). We would also like to express our gratitude to the Tianshan Youth Project of Xinjiang No.2017Q014.

\software{MESA(v8845; Paxton et al. 2011, 2013, 2015), BSE (Hurley et al. 2000, 2002)}



\bibliographystyle{./aasjournal}
\bibliography{./sample62}

\include{table_information}
\include{table_KS}
\include{figure}

\end{document}